\documentclass[twocolumn,showpacs,prl,aps]{revtex4}

\usepackage{epsfig}
\usepackage{amsmath}

\begin{document}

\title{Topological disentangler for the valence-bond-solid chain}
\author{Kouichi Okunishi}
\affiliation{Department of Physics, Faculty of Science, Niigata University, Niigata 950
-2181, Japan.}

\date{\today}
\begin{abstract}
We discuss topological disentangler for  $S=1$ quantum spin chains in the Haldane phase.
We first point out that Kennedy-Tasaki's(KT) nonlocal unitary transformation is the perfect disentangler for Affleck-Kennedy-Lieb-Tasaki model. 
We then demonstrate that the KT transformation can be reconstructed as an assembly of pair disentanglers.
Finally, we show that the KT transformation can be regarded as a topological disentangler, which selectively disentangles the double-fold degeneracy in the entanglement spectrum of the $S=1$ Heisenberg chain.
\end{abstract}

\pacs{75.10.Kt,03.65.Ud}

\maketitle


Topological aspects of  low-dimensional quantum many body systems have been attracting considerable interest in modern physics.
One of the most fundamental examples is  the Haldane-gap system for $S=$integer spin chains.\cite{haldane}
According to the continuous efforts since the Haldane's conjecture based on the non-linear sigma model,  a couple of interesting concepts in the low-dimensional physics have been developed:   valence-bond-solid(VBS) states and  effective $S=1/2$ edge spins\cite{AKLT}, topological string order\cite{dennijis}, spontaneous breaking of the hidden $Z_2\times Z_2$ symmetry\cite{KT},  $Z_2$ Berry phase\cite{hatsugai}, etc.
Recently, the topological order in quantum spin systems has been illuminated by the entanglement spectrum. 
It is shown that the non-trivial degeneracy appears in the entanglement spectrum\cite{pollmann}, which is closely related to the topological order protected by the symmetry\cite{gu}.
This suggests that the connection between the entanglement and the topological order becomes important.

The entanglement of the groun dstate wavefunction also provides an indispensable view point for numerical renormalization groups of quantum many body systems.
It is well established that density matrix renormalization group(DMRG)\cite{white} is a variational method for the matrix product(MP) type wavefunction\cite{ostlund}, which maximizes the block entanglement entropy.
Recently, multi-scale entanglement renormalization(MERA) is proposed to be a powerful numerical simulation method\cite{mera}.
A key point in MERA is that the entanglement entropy is reduced by the combination of the usual block-spin transformation and the local unitary transformation called ``disentangler" that disentangles the quantum entanglement of the neighboring effective spins.
Successive operations of the disentanglers and  the block-spin transformations drastically improve accuracy in contrast to the conventional renormalization group.
This suggests that the disentangler is deeply related to the structure of the ground state wavefunction.
In this sense, to understand how the disentangler controls the entanglement is an essential problem in physics of quantum many body systems.

In order to discuss the connection of the disentangler and topological order,  the most striking play ground is  Affleck-Kennedy-Lieb-Tasaki(AKLT) model, whose ground state is exactly described by the VBS state\cite{AKLT}.
Let us recall that  the Kennedy-Tasaki's(KT) non-local unitary transformation plays a central role to clarify the topological order and the hidden $Z_2\times Z_2$ symmetry in the Haldane phase\cite{KT}.
The KT transformation converts the nonlocal string order parameter into the {\ classical} ferromagnetic order parameter with the manifest  $Z_2\times Z_2$ symmetry.
In this letter, we point out that the KT transformation is nothing but the perfect disentangler of the VBS state.
We then demonstrate that the KT transformation can be reconstructed  by the pair disentanglers, which disentangle any pair of spins in the AKLT chain. 
In addition, for the case of $S=1$ Heisenberg chain, the KT transformation particularly disentangles the two-fold degeneracy of the entanglement spectrum originating from the topological order, which implies that the KT transformation can be interpreted as a topological disentangler.

The Hamiltonian of the AKLT model is given by ${\cal H}_{\rm AKLT} \equiv \sum_ih_{i,i+1} $ with
\begin{equation}
h_{i,i+1}= \vec{S}_i\cdot \vec{S}_{i+1} +\frac{1}{3}(\vec{S}_i\cdot \vec{S}_{i+1})^2,
\end{equation}
where $\vec{S}$ represent the $S=1$ spin matrices.
The KT transformation, which is denoted as $\cal U$, leads 
${\cal U} \vec{S}_i\cdot \vec{S}_{i+1} {\cal U}^{-1}= 
-\tilde{S}_i^x\tilde{S}^x_{i+1}-\tilde{S}_i^y\tilde{S}^y_{i+1} +\tilde{S}_i^xe^{i\pi (S_i^z +S_{i+1}^x)}\tilde{S}^x_{i+1}
$.
For the AKLT Hamiltonian transformed by the KT transformation, the infinite-volume ground state is calculated by diagonalizing the local Hamiltonian $\tilde{ h}_{i,i+1}={\cal U} {h}_{i,i+1} {\cal U}^{-1}$,  
\begin{equation}
|\Phi^{\nu}\rangle = \cdots |\phi^\nu\rangle \otimes|\phi^\nu\rangle\otimes|\phi^\nu\rangle\otimes \cdots, ~  {\rm for}~ \nu=1,2,3,4. \label{ktwf}
\end{equation}
where 
\begin{eqnarray}
|\phi^{1}\rangle &=&  \sqrt{2/3} |+\rangle  +  \sqrt{1/3} |0\rangle, \nonumber \\
|\phi^{2}\rangle &=&  \sqrt{2/3} |+\rangle  -  \sqrt{1/3} |0\rangle, \nonumber  \\
|\phi^{3}\rangle &=&  \sqrt{2/3} |-\rangle  +  \sqrt{1/3} |0\rangle, \nonumber \\
|\phi^{4}\rangle &=&  \sqrt{2/3} |-\rangle  -  \sqrt{1/3} |0\rangle.  
\end{eqnarray}
Since these four degenerating states $|\Phi^\nu\rangle$  are represented as the direct products of $|\phi^\nu\rangle$, there is no correlation between different sites and thus the entanglement entropy for any block size is exactly zero.
This implies that {\it the KT transformation works as the perfect disentangler of the VBS state} with an appropriate boundary condition.

The KT transformation is originally introduced as a sequential operation for a spin alignment along the chain.\cite{KT}
Here, let us assign the site index 1 to $N$ from left to right along the chain, where $N$ is an even integer representing the length of the chain.
If the number of ``$+$'' and ``$-$'' spins sitting in the left of a certain site $j$ is odd, then the spin at the $j$-th site is flipped. 
In addition, if the total number of ``0'' spins at the odd sites in the entire chain is odd, then  a minus sing is assigned to the state vector.
This is a complicated operation for the Hilbert space, but the explicit from of the KT transformation can be written as 
\begin{equation}
{\cal U} = \prod_{l=1}^N\prod_{k=1}^{l-1} D_{k,l}
\label{kttransform}
\end{equation}
where $D_{k,l} =  D^{-1}_{k,l} \equiv e^{i\pi S_k^z \otimes S_l^x }$ and the overall sing is omitted.\cite{oshikawa}
Here, we also present another form of $D_{k,l}$, which is  convenient in practical calculations,
\begin{eqnarray}
D_{k,l}      & =& P^{\pm}_k\otimes e^{i\pi S^x_l}+P^{0}_k\otimes \openone_l \nonumber \\
      & =&  e^{i\pi S^z_k}\otimes Q^{\pm}_l+\openone_k\otimes Q^{0}_l \label{doperator}
\end{eqnarray}
where $\openone_k$ is the $3\times 3$ identity matrix for $S=1$ spin at $k$th site and $ P^{\pm}_k$($ P^{0}_k$) is the projection operator into the $S^z=\pm 1$($S^z=0$) space at $k$th site.
Explicitly, we have $ P^{\pm}_l=\frac{1}{2}(\openone_k-e^{i\pi S_k^z})$ and $ P^{0}_k=\frac{1}{2}(\openone_k+e^{i\pi S_k^z})$.
Similarly, $ Q^{\pm}_l$($ Q^{0}_l$) is the projection operator into $S^x=\pm 1$($S^x=0$) space in the $S^x$-diagonal representation.
As was pointed out in Ref.\cite{oshikawa},
$[D_{k,l}, D_{k',l'} ] =0$ for $k<l$ and $k'<l'$, implying that  the order of the $D$ operators is not relevant in Eq. (\ref{kttransform}).
Thus the KT transformation can be constructed as the assembly of $D$ for the all spin pairs, as is depicted in Fig. 1.
This suggests that the entanglement of any spin pair in the VBS state is disentangled by $D$.
We thus call  $D$ ``pair disentangler'' in the following.

\begin{figure}[t]
\epsfig{file=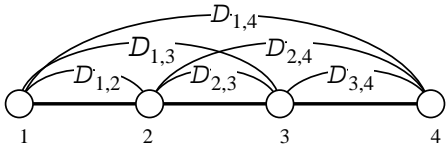,width=6cm}
\caption{Schematic diagram of the KT transformation for a 4 spin system.
The KT transformation is represented as a assembly of the disentanglers for all pairs}
\end{figure}

Let us discuss  disentangling the VBS state by the pair disentanglers.
The degenerating ground state (\ref{ktwf}) was originally obtained by diagonalizing the KT-transformed Hamiltonian $\tilde{h}_{i,i+1}$. 
Here, we directly investigate how the pair disentangler disentangles the VBS state.
The VBS state for the AKLT model can be compactly written in the MP form,\cite{MPAKLT}
\begin{equation}
{\Psi} = A_1 A_2 \cdots A_N \Omega \label{mpstate},
\end{equation}
where ${\Psi}$ is the 2$\times$2 matrix  and the four entries correspond to the four degenerating eigenstate of the AKLT Hamiltonian with the open boundary condition.
The explicit form of the matrix $A$ is given by
\begin{equation}
A_i = \begin{pmatrix}
-\sqrt{\frac{1}{3}} |0\rangle_i &\sqrt{\frac{2}{3}} |+\rangle_i \\
-\sqrt{\frac{2}{3}} |-\rangle_i & \sqrt{\frac{1}{3}} |0\rangle_i 
\end{pmatrix}
\end{equation}
where the kets represent the $S^z$-diagonal bases of the $S=1$ spin at site $i$. 
Here, it should be noted that these states are the non-orthogonal basis of a finite size system and the orthogonality  is recovered in the infinite size limit.
In Eq. (\ref{mpstate}), we have also introduced the boundary matrix
\begin{equation}
\Omega \equiv \begin{pmatrix} 
 1 & 1 \\
-1 & 1
\end{pmatrix}
,
\end{equation}
which just yields a linear combination of the four degenerating VBS states.
A possible physical interpretation of $\Omega$ is as follows.
According to Eq. (\ref{doperator}), we can see that the spins in the left side of the pair disentangler are written in the usual $S^z$-diagonal representation, while the $S^x$-diagonal basis is rather natural for the spins in the right side of the pair disentangler.
In order to treat the both right and left edges equivalently in the pair disentangler, it is appropriate to take the $S^x$-diagonal representation of the spin at the right edge.
Then, for $\pi/2$-rotation around the $y$ axis,  we find 
\begin{equation}
e^{i \pi S^y /2} A = \Omega^{-1} A \Omega.
\end{equation}
This suggests that the boundary matrix $\Omega$  adjusts the quantization axis at the right edge and  a certain domain wall is possibly inserted in the MP state of the bulk region. 

\begin{figure}[t]
\epsfig{file=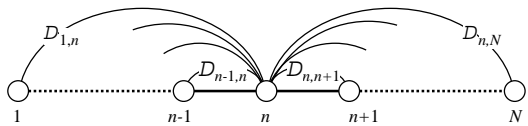, width=7cm}
\caption{Disentangling the spin at  $n$th site.}
\end{figure}

We now demonstrate that a certain $n$th site in the VBS state can be disentangled by the disentanglers depicted in Fig. 2. 
We operate the pair disentanglers between the $n$th site and the other sites to the MP wavefunction,
\begin{equation}
\tilde{\Psi}= \prod_{j=1}^{n-1} D_{j,n} \prod_{j=n+1}^{N} D_{n,j} {\Psi} .
\label{singlespinD}
\end{equation}
Using Eq. (\ref{doperator}), we have
\begin{eqnarray}
\prod_{j=1}^{n-1} D_{j,n}&=& e^{i\pi S_1^z} \otimes e^{i\pi S_{2}^z} \otimes \cdots e^{i\pi S_{n-1}^z}  \otimes Q^{\pm}_n \nonumber \\
&+& \openone_1 \otimes \openone_{2}\cdots \otimes \openone_{n-1} \otimes  Q^{0}_n ,\\
\prod_{j=n+1}^{N} D_{n,j}&=&P^{\pm}_k\otimes e^{i\pi S^x_{n+1} }\otimes e^{i\pi S^x_{n+2} }\cdots \otimes e^{i\pi S^x_{N} }\nonumber \\
&+&P_n^0\otimes \openone_{n+1} \otimes \openone_{n+2} \cdots \otimes \openone_N.
\end{eqnarray}
When applying these operators to the MP state, the following relations are useful, 
\begin{equation}
e^{i \pi S^z } A = \sigma^z A \sigma^z, \qquad e^{i \pi S^x } A = \sigma^x A \sigma^x,
\label{gauge1}
\end{equation}
where $\sigma^z$ and $\sigma^x$ are the Pauli matrices acting in the axially space. 
Since the adjacent Pauli matrices in the bulk part can be canceled with each other, $\sigma^{x,z}$ emerges at the boundaries.
Then, Eq. (\ref{singlespinD}) becomes
\begin{eqnarray}
&& \sigma^zA_1\cdots A_{n-1}\sigma^z (P_n^\pm Q_n^\pm A_n )\sigma^x A_{n+1}\cdots A_N\sigma^x \Omega \nonumber\\
&&+ \sigma^zA_1\cdots A_{n-1}\sigma^z( P_n^0 Q_n^\pm A_n )A_{n+1}\cdots A_N \Omega \nonumber\\
&&+ A_1\cdots A_{n-1}( P_n^\pm Q_n^0 A_n )\sigma^xA_{n+1}\cdots A_N\sigma^x \Omega \nonumber\\
&&+ A_1\cdots A_{n-1}(P_n^0 Q_n^0 A_n )A_{n+1}\cdots A_N \Omega.
\label{gauge2}
\end{eqnarray}
Here, we introduce the notation,
\begin{eqnarray*}
A_1\cdots A_{n-1} = 
\begin{pmatrix}
 |\alpha_1 \rangle &  |\beta_1\rangle \\
 |\gamma_1 \rangle &  |\delta_1\rangle
\end{pmatrix}, \\
A_{n+1}\cdots A_{N-1} = 
\begin{pmatrix}
 |\alpha_2 \rangle &  |\beta_2\rangle \\
 |\gamma_2 \rangle &  |\delta_2\rangle
\end{pmatrix}.
\end{eqnarray*}
These matrix elements are complex linear combinations of the $S=1$-kets for $1\cdots n-1$ or $n+1\cdots N$ sites. But we do not need the explicit form below.
A straightforward calculation yields
\begin{equation}
\tilde{\Psi} =
\begin{pmatrix}
|\phi^1\rangle \otimes |X_{11}\rangle & 
|\phi^2 \rangle \otimes |X_{12}\rangle  
\\
|\phi^4\rangle \otimes |X_{21}\rangle  & |\phi^3\rangle \otimes |X_{22}\rangle 
\end{pmatrix}
,
\label{spindisentangled}
\end{equation}
where  $|X_{lm}\rangle$ with $l,m = 1,2$ are also complex linear combinations of the $S=1$ kets corresponding to  $j=1\cdots n-1, n+1 \cdots N$ spins.
They are explicitly given by 
\begin{eqnarray}
|X_{11}\rangle&=&
|\alpha_1\rangle |\beta_2 \rangle +|\beta_1\rangle  |\delta_2\rangle -|\alpha_1\rangle |\alpha_2 \rangle - |\beta_1\rangle |\gamma_2 \rangle , \nonumber \\
|X_{12}\rangle &=&
 |\alpha_1\rangle |\beta_2 \rangle +|\beta_1\rangle |\delta_2\rangle +|\alpha_1\rangle  |\alpha_2 \rangle + |\beta_1\rangle|\gamma_2 \rangle  , \nonumber \\
|X_{21} \rangle& =&
 |\gamma_1\rangle |\alpha_2\rangle + |\delta_1 \rangle |\gamma_2 \rangle- |\gamma_1\rangle |\beta_2\rangle -|\delta_1 \rangle |\delta_2 \rangle ,  \\
|X_{22} \rangle &=&
- |\gamma_1\rangle |\alpha_2\rangle - |\delta_1 \rangle |\gamma_2 \rangle  -|\gamma_1\rangle |\beta_2\rangle -|\delta_1 \rangle |\delta_2 \rangle ,  \nonumber  \label{elements}
\end{eqnarray}
where the symbol of tensor product is omitted for simplicity.
A significant point of Eq. (\ref{spindisentangled}) is that, in each of the four matrix elements,  $|\phi^\nu\rangle$ of the $n$th site targeted is clearly decoupled from $|X_{lm}\rangle$ by the direct tensor product.
Thus we can verify that the spin at the $n$th site can be disentangled with the other spins in the chain.

We construct the single-spin density matrix at the $n$th site for $|\phi^1\rangle \otimes |X_{11}\rangle$ or $|\phi^2 \rangle \otimes |X_{12}\rangle$. The result is easily obtained as 
\begin{equation}
\tilde{\rho}= \begin{pmatrix}
\frac{2}{3} & \frac{\sqrt{2}}{3} & 0 \\
 \frac{\sqrt{2}}{3} & \frac{1}{3} & 0 \\
0 & 0& 0
\end{pmatrix} .
\end{equation}
The entanglement spectrum is  clearly $(1, 0, 0)$ and its entanglement entropy is ${\cal S}= 0$.
Also we have the similar results for $|\phi^4\rangle \otimes |X_{21}\rangle$ and $ |\phi^3\rangle \otimes |X_{22}\rangle $.
This result should be contrasted to the entanglement spectrum of the original VBS state.
The single-spin density matrix of the VBS state is obtained as $\rho= \sum_{m=+,0,-} \frac{1}{3}|m\rangle\langle m|$ with the appropriate edge spins and then the entanglement entropy is ${\cal S}=\ln 3$,  implying that the single spin in the VBS state is  maximally entangled with the other spins.\cite{korepin}
Thus our disentangler completely disentangles the spin at $n$th site from the other spins.

A generalization to the general block entanglement is straightforward. 
When we disentangle a system block of a finite length from the other part of the chain,  we should construct a couple of the pair disentanglers for the all pairs linking the system and the bath. 
Moreover,  recursively disentangling the spins from left to right in the similar manner to Eq. (\ref{spindisentangled}), we can finally reproduce Eq. (\ref{ktwf}).

We turn to the $S=1$ Heisenberg chain, which is described by ${\cal H}\equiv \sum_i \vec{S}_i\cdot \vec{S}_{i+1}$.
Although the ground state of the Heisenberg chain is adiabatically connected to the VBS state, it can not be expressed by the MP state with a finite dimension.
What happens on the entanglement spectrum of the KT-transformed Heisenberg model $\tilde{\cal H} ={\cal U H U}^{-1} $ ?
In order to discuss the relation between the KT transformation and the entanglement spectrum for the Heisenberg case, we employ the  product-wavefunction renormalization group\cite{pwf}, which is a variant of the infinite-system-size DMRG and enables for us to directory deal with the bulk limit. 
We then evaluate the eigenvalue spectrum of the reduced density matrix for the half-infinite chain.
Note that, as was mentioned in Ref. \cite{KT},  the SU(2) symmetry is masked by the KT transformation and the total-$S^z$ conservation is not available in DMRG computation of ${\tilde{\cal H}}$.

\begin{figure}[t]
\epsfig{file=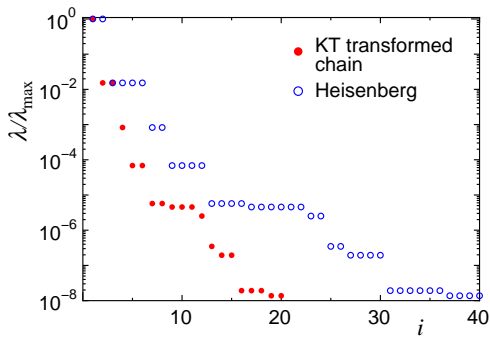,width=6.5cm}
\caption{Entanglement spectrum of the half-infinite reduced density matrix for the $S=1$ Heisenberg chain(open circles) and the Kennedy-Tasaki transformed chain(solid circles). The spectrum is normalized by the largest eigenvalue.}
\label{fig3}
\end{figure}

The eigenvalues spectrum $\lambda$ of the half-infinite density matrix  is shown in Fig.\ref{fig3}, where the retained number of basis in the DMRG computation is $m=200$ and the spectrum is normalized by  the largest eigenvalue $\lambda_{\rm max}$.
The spectrum of the Heisenberg model shows the double-fold degeneracy, which reflects the topological $Z_2$ symmetry.
On the other hand, the spectrum for $\tilde{\cal H}$ has no double-fold degeneracy.
Then, an interesting point is that the spectra of $\tilde{\cal H}$ and ${\cal H}$ are identical to each other except for the double-fold degeneracy;
We have confirmed $\tilde{\lambda}_i=\lambda_{2i}$ within the numerical accuracy, where $\tilde{\lambda}_i$ and $\lambda_{i}$ respectively denote the entanglement spectra for $\tilde{\cal H}$ and ${\cal H}$.
As a result, we note that the variational calculation for $\tilde{\cal H}$ in Ref. \cite{KT} is eventually equivalent to that of the MP state for ${\cal H}$ with the matrix size $m=4$, which is also the same as a DMRG computation with $m=4$.

In the context of physics, the topological $Z_2\times Z_2$ symmetry is spontaneously broken in  $\tilde{\cal H}$ and one of the four degenerating state is selected.
The KT transformation affects only the degeneracy in the entanglement spectrum  originating from the $Z_2$-edge spin.
Accordingly, the entanglement entropy of $\tilde{\cal H}$ is reduced from that of ${\cal H}$ by $\tilde{\cal S}={\cal S} -\ln 2$, corresponding to the topological entanglement entropy\cite{kitaev}.
We therefore conclude that the KT transformation and the pair-disentangler (\ref{doperator}) play the role of {\it the topological disentangler}.
In other words,  the disentangler for the topological symmetry is distinguishable from the disentangler associated with the entanglement spectrum of the other dynamical behaviors.

To summarize, we have constructed the exact topological disentangler for the arbitrary block of the Haldane-gap chains by combining the pair disentaglers.
Then, the important point is that the KT transformation, which is nothing but the disentangler of the entire system,  can be reconstructed by the pair disentanglers for the all spin pairs. 
The exact topological disentangler involves a couple of implications for physics in the low-dimension.
The Heisenberg model results indicates that the entanglement due to the topological symmetry can be distinguished from that of the other dynamical origin. This suggests that, through the disentangler, the topological symmetry can be priori taken into account in the MP formulation including numerical computation.
Next, although construction of a non-local transformation  has been a highly non-trivial problem, the decomposition and reconstruction of the pair disentanglers provide a systematical approach to find the non-local transformation for the general topological orders.
Indeed,  the generalized string order for the higher-$S$ spin chain are actually obtained in Ref.\cite{oshikawa}.
This implies that the topological disentangler can be straightforwardly constructed for a class of the VBS states.
We further mention that the exact disentangler is also important from the numerical-simulation view point.
For example, the global entanglement in MERA is reduced by the layered structure of the tensor network state, and the disentanglers  are obtained as numerics after several optimization process.
Then, the exact disentangler is of great use to check the quality of the disentangler in  numerical simulations.
We finally point out an interesting connection to the MP formulation of Bethe ansatz.
In the present construction of the disentangler,   the boundary matrix $\Omega$  appears. 
In a recent study of the MP Bethe ansatz, the very similar domain-wall boundary matrix also appears, where the gauge transformation like Eqs. (\ref{gauge1}) and (\ref{gauge2}) plays a crucial role.\cite{katsura}
The construction of the disentangler for the integrable system  may be an essential future problem.
We believe that the exact disentangler develops various frontiers of  quantum many-body physics.


The author thanks T. Nishino, I. Maruyama  and G. Vidal for valuable discussions.
This work is supported by Grant-in-Aid for Scientific Research from Ministry of Education and Science Japan(No.20340096).



\begin{thebibliography}{99}


\bibitem{haldane} F. D. M. Haldane, Phys. Lett. A {\bf 93}, 464 (1983); Phys. Rev. Lett. {\bf 50}, 1153 (1983).


\bibitem{AKLT} I. Affleck, T. Kennedy, E. H. Lieb, and H. Tasaki, Phys. Rev. Lett. {\bf 59}, 799 (1987). 

\bibitem{dennijis} M. den Nijs and K. Rommelse, Phys. Rev. B {\bf 40}, 4709 (1989).

\bibitem{KT} T. Kennedy and H. Tasaki, Phys. Rev. B {\bf 45}, 304 (1992).

\bibitem{hatsugai} T. Hirano, H. Katsura, and Y. Hatsugai, Phys. Rev. B {\bf 77}, 094431 (2008).

\bibitem{pollmann}F. Pollmann, A. M. Turner, E. Berg, and M. Oshikawa, Phys. Rev. B {\bf 81}, 064439 (2010).

\bibitem{gu} Z.-C. Gu and X.-G. Wen, Phys. Rev. B 80, 155131 (2009). 

\bibitem{white} S.R. White, Phys. Rev. Lett. {\bf 69}, 2863 (1992).

\bibitem{ostlund} S. \"Ostlund and S. Rommer,  Phys. Rev. Lett. {\bf 75}, 3537 (1995).

\bibitem{mera} G. Vidal, Phys. Rev. Lett. {\bf 99}, 220405 (2007); G. Evenbly and G. Vidal. Phys. Rev. B {\bf 79}, 144108 (2009).


\bibitem{oshikawa} M. Oshikawa, J. Phys.: condens.matt. {\bf 4}, 7469 (1992).


\bibitem{MPAKLT} M. Fannes, B. Nachtergaele, and R. W. Werner, Commun. Math. Phys. {\bf 144}, 443 (1992). 
A.Kl\"umper, A.  Schadschneider, and J. Zitterz, Eur. Phys. Lett. {\bf 24}, 293 (1993).

\bibitem{korepin} H. Fan, V. Korepin and V.  Roychowdhury, Phys. Rev. Lett. {\bf 93}, 227203 (2004).

\bibitem{pwf} T. Nishino and K. Okunishi, J. Phys. Soc. Jpn. {\bf 64}, 4084 (1995).
Y. Hieida, K. Okunishi and Y. Akutsu, Phys. Lett. A {\bf 233}, 464 (1997). 

\bibitem{kitaev} A. Kitaev and J. Preskill, Phys. Rev. Lett. {\bf 96}, 110404 (2006)

\bibitem{katsura} H. Katsura and I. Maruyama, J. Phys.: Math. Gen. {\bf 43}, 175003 (2010).



\end{thebibliography}
\end{document}